\documentclass{pasj00}

\begin{document}
\SetRunningHead{Goto et al.           }{H$_3^+$ toward The Galactic Center}

\Received{2002/08/23}
\Accepted{2002/10/07}

\title{Absorption Line Survey of H$_3^+$
toward the~Galactic~Center~Sources I.
GCS~3-2~and~GC~IRS3 \footnotemark[\ast] }




\author{
  Miwa        \textsc{Goto        },\altaffilmark{1,2}
  Benjamin J. \textsc{McCall      },\altaffilmark{3}
  Thomas R.   \textsc{Geballe     },\altaffilmark{4}
  Tomonori    \textsc{Usuda       },\altaffilmark{1}\\
  Naoto       \textsc{Kobayashi   },\altaffilmark{1}
  Hiroshi     \textsc{Terada      },\altaffilmark{1}
  and
  Takeshi     \textsc{Oka         }\altaffilmark{5}}

\altaffiltext{1}{Subaru Telescope, 650, North A'ohoku Place,
                 Hilo, HI 96720, USA}
\email{mgoto@naoj.org,mgoto@duke.ifa.hawaii.edu}

\altaffiltext{2}{Institute for Astronomy, University of Hawaii,
                 640, North A'ohoku Place,  Hilo, HI 96720, USA}

\altaffiltext{3}{Department of Chemistry and Department of Astronomy,\\
                 University of California, 601 Campbell Hall,
                 Berkeley, CA 94720-3411, USA}

\altaffiltext{4}{Gemini Observatory,
                 670, North A'ohoku Place,  Hilo, HI 96720, USA}

\altaffiltext{5}{Department of Astronomy and Astrophysics,
                 Department of Chemistry, and Enrico Fermi Institute,\\
                 University of Chicago, Chicago, IL 60637, USA}



\maketitle

\footnotetext[$\ast$]{Based on data collected at Subaru Telescope,
which is operated by the National Astronomical Observatory of Japan.}

\begin{abstract}

We present high-resolution ($R$ = 20000) spectroscopy of H$_3^+$
absorption toward the luminous Galactic center sources GCS~3-2 and GC
IRS~3.  With the efficient wavelength coverage afforded by Subaru
IRCS, six absorption lines of H$_3^+$ have been detected in each
source from 3.5 to 4.0~$\mu$m, three of which are new. In particular
the 3.543~$\mu$m absorption line of the $R$(3, 3)$^l$ transition
arising from the metastable ($J$, $K$) = (3, 3) state has been
tentatively detected for the first time in the interstellar medium,
where previous observations of H$_3^+$ had been limited to absorption
lines from the lowest levels: ($J$, $K$) = (1, 0) of ortho-H$_3^+$ and
(1, 1) of para-H$_3^+$.

The H$_3^+$ absorption toward the Galactic center takes place in dense
and diffuse clouds along the line of sight as well as the molecular
complex close to the Galactic nucleus. At least four kinematic
components are found in the H$_3^+$ absorption lines. We suggest
identifications of the velocity components with those of H {\sc i},
CO, and H$_2$CO previously reported from radio and infrared
observations.  H$_3^+$ components with velocities that match those of
weak and sharp CO and H$_2$CO lines are attributed to diffuse
clouds. Our observation has revealed a striking difference between the
absorption profiles of H$_3^+$ and CO, demonstrating that the
spectroscopy of H$_3^+$ provides information complementary to that
obtained from CO spectroscopy.

The tentative detection of the $R$(3, 3)$^l$ line and the
non-detection of spectral lines from other $J > 1$ levels provide
observational evidence for the metastability of the (3, 3) level,
which is theoretically expected. This suggests that other
metastable $J$ = $K$ levels with higher $J$ may also be populated.

\end{abstract}

\section{Introduction}
The crucial role which the H$_3^+$ molecular ion plays in interstellar
chemistry was first addressed by \citet{wat73} and
\citet{her73}. H$_3^+$ is produced by cosmic-ray ionization of H$_2$
to H$_3^+$ followed by the efficient Langevin reaction,
\begin{equation}
{\rm H_2^+ + H_2 \rightarrow H_3^+ + H.}
\end{equation}
 It works as a universal protonator (acid) in the efficient proton hop
 reaction
\begin{equation}
{\rm H_3^+ + X \rightarrow H_2 + XH^+ }
\end{equation}
for most molecules or atoms X (He, Ne, N, and O$_2$ are a few
exceptions). Subsequent general reactions with a molecule {\rm Y},
\begin{equation}
{\rm XH^+ + Y \rightarrow XY^+ + H},
\end{equation}
lead further to chemical networks which produce complex molecules
in dense interstellar clouds.

\newpage
A search for interstellar H$_3^+$ was initiated by \citet{oka81} more
than two decades ago, and culminated in the discovery of H$_3^+$ in
the interstellar medium toward AFGL~2136 and W33A, young stellar
objects deeply embedded in molecular clouds \citep{geb96}. Since then
the study of interstellar H$_3^+$ has progressed rapidly by a series
of successful observations in dense clouds \citep{mcc99} and in
diffuse clouds toward the highly reddened star Cygnus OB2 12, the
Galactic center, and many other sightlines \citep{mcc98b, geb99,
mcc02}. These observations have not only demonstrated the ubiquity and
abundance of this molecular ion, which was anticipated in general in
molecular hydrogen dominated plasmas \citep{mar61}, but also revealed
an interesting enigma related to the H$_3^+$ chemistry of the diffuse
interstellar medium. While the H$_3^+$ column densities measured in
dense clouds agreed well with those predicted by the H$_3^+$
production and destruction mechanisms of \citet{her73} \citep{geb96,
mcc99}, those measured in the diffuse interstellar medium have been
found to be orders of magnitude higher than estimated using canonical
values of plasma chemical constants (McCall et al. 1998a,b; Geballe et
al. 1999, McCall et al. 2002). 
In order to have a better understanding of this problem, we undertook
an absorption-line survey of this molecular ion toward the Galactic
center.  This study also provides unique information on the medium
toward the Galactic center because of the special characteristics of
H$_3^+$ as an astrophysical probe.

The absorption-line survey was designed to cover the maximum number of
H$_3^+$ absorption lines in clouds with various physical
conditions. The Galactic center sources are ideal for this purpose,
since they suffer heavy visual extinction of $A_{V}$ = 25--40
\citep{cot00}, the highest in the Galaxy among those obscured mainly
by diffuse clouds. The line of sight to the Galactic center cuts
through dense and diffuse clouds in the intervening spiral arms as
well as the molecular complex close to the nucleus.  Thus, H$_3^+$ in
different physical conditions is efficiently sampled in one
observation.  The intervening clouds have different radial velocities,
and the nature of the clouds were studied in previous spectroscopic
observations using molecular and atomic tracers from near-infrared to
radio wavelengths. Two infrared sources, GCS~3-2 in the Quintuplet
cluster \citep{nag90,oku90} and GC~IRS~3 \citep{bec75, bec78} near Sgr
A$^\ast$ have been selected, since both appear to have intrinsically
featureless spectra due to dust emission and are sufficiently luminous
to provide good continuum fluxes for $L$-band absorption spectroscopy.

Wide wavelength coverage is essential for a line survey. The
observation of interstellar H$_3^+$ has so far been almost entirely
limited to the three absorption lines, [$R$(1, 0), $R$(1, 1)$^l$, and
$R$(1, 1)$^u$], all arising from the lowest $J$ = 1 rotational levels,
i.e.\ ($J$, $K$) = (1, 0) of ortho-H$_3^+$ and (1, 1) of para-H$_3^+$
(note that the $J$ = 0 level is forbidden by the Pauli principle;
McCall 2001). This is because high spectral resolution and wide
wavelength coverage are often incompatible. This limitation has now
been overcome by a new generation of cross-dispersed infrared
spectrographs, such as IRCS on the Subaru Telescope.  We can now
simultaneously observe at high spectral resolution not only most of
the transitions starting from the $J$ = 1 levels, including $Q$(1, 0)
and $Q$(1, 1), but also other transitions starting from higher $J$
levels. This wide coverage has led us to a tentative detection of the
$R$(3, 3)$^l$ line starting from the metastable ($J$, $K$) = (3, 3)
level.

\section{Observations}
These spectroscopic observations were made on 2001 June 16 (UT) using
the Infrared Camera and Spectrograph (IRCS; Tokunaga et al. 1998;
Kobayashi et al. 2000) with the 8.2~m Subaru Telescope on Mauna
Kea. Subaru IRCS is equipped with an echelle and a cross-dispersing
grating to offer high resolution spectroscopy without sacrificing
wavelength coverage. Figure \ref{fig:cov1} shows the wavelength
coverage of the IRCS echelle mode in the 3~$\mu$m region. The entire
$L$-band (2.84--4.18 ~$\mu$m) can be covered with six standard
settings, with 70\% coverage in three settings. Thus, Subaru IRCS is a
good match for a survey of lines scattered in a single band. The
grating setting can be modified continuously for the best coverage of
target lines in order to maximize observing efficiency.

In the current experiment we employed two grating settings optimized
for coverage of the $\nu_2 \leftarrow 0$ H$_3^+$ lines in the 3~$\mu$m
region illustrated in figure \ref{fig:cov1}. At the time of the
observations, there was high dark current in the upper two quadrants
of the 1k $\times$ 1k InSb detector array, so we covered the relevant
wavelength ranges using only the lower half of the array.  
A 0\farcs15 $\times$ 4\farcs5 slit was used to achieve $R$ = 20000
spectral resolution. The spectra were recorded by nodding the
telescope between two points separated by 2\farcs2 along the slit to
subtract the sky emission and dark current.  Spectroscopic standard
stars were observed through airmasses similar to those of the objects
in order to cancel out the atmospheric transmission efficiency.  The
details of the observations are summarized in table
\ref{tab:log1}. The seeing was not better than 0\farcs7 at $L'$, and
varied during the observing period.  Spectroscopic flat frames were
obtained at the end of the night with a halogen lamp in the telescope
calibration unit installed in front of the instrument window.

\section{Data Reduction}

The observed spectrograph images were stacked and averaged without any
pixel registration since both the accumulation of the instrument
flexure and the telescope tracking error during automated guiding was
negligible over the typical tracking period. The average frame was
flat-fielded with a dark-subtracted halogen lamp frame. Bad pixel
masks were created by collecting pixels with low, high, or varying
responses, or with a high dark current, based on the statistics of the
halogen lamp and the dark-current frames. Those outlier pixels were
filtered out before spectral extraction.

One dimensional spectra were obtained using the IRAF\footnote{IRAF is
distributed by the National Optical Astronomy Observatories, which are
operated by the Association of Universities for Research in Astronomy,
Inc., under cooperative agreement with the National Science
Foundation.} aperture-extraction package. Spectroscopy in the 3~$\mu$m
region suffers severely from interference by the absorption lines of
molecules in the Earth's atmosphere. Poor cancellation of the
atmospheric lines by way of dividing by the spectra of the standard
stars is often the primary source of systematic errors, which easily
overwhelm statistical noise. We carefully examined the locations of
the H$_3^+$ lines listed in table \ref{tab:lin1} using a custom
written IDL code. The code handles (1) linear registration of the
wavelength offset between the object and the spectroscopic standard,
(2) rescaling the normalized standard star spectrum according to
Beer's law to minimize the airmass mismatch that sometimes exceeded
10\% in our observation, and (3) correction of the spectral resolution
by deconvolving the spectra.
 Wavelength calibration was performed by maximizing the cross
correlation between the object spectrum and the model atmospheric
spectrum calculated by ATRAN code \citep{lor92}. 

\section{Results}
The result of the line survey is shown in figure \ref{fig:spe1} and
figure \ref{fig:neg2} for the positive and negative detections,
respectively. The model atmospheric transmission curves convolved to
the same spectral resolution are shown in order to discriminate
between the real spectral features and the residual features of the
strong telluric absorption lines. All of the absorption lines in our
coverage starting from the lowest $J$ = 1 levels of ortho- and
para-H$_3^+$ [$R$(1, 0), $Q$(1, 0), $R$(1, 1)$^l$, $R$(1, 1)$^u$, and
$Q$(1, 1)] were successfully detected, while those from $J > 1$ levels
were all negative, except for $R$(3, 3)$^l$.

The H$_3^+$ absorption lines from the $J$ = 1 levels show several
discrete kinematic components. The absorption profiles are deconvolved
into Gaussian components centered at LSR velocities of $-$110 --
$-$140~km~s$^{-1}$, $-$50 -- $-$60~km~s$^{-1}$, and 0~km~s$^{-1}$
(figure. \ref{fig:dec1}). The deconvolution parameters are summarized
in table \ref{tab:vel1}.  GC~IRS~3 shows wide absorption at
$-$20~km~s$^{-1}$ and a wing at +50~km~s$^{-1}$ on a broad pedestal
absorption. A weak absorption is seen at $-$170~km~s$^{-1}$ in
GCS~3-2. Note that the velocity components are less distinguishable at
3.668~$\mu$m, previously observed toward the Galactic center
\citep{geb99}. This is because the absorption feature consists of a
close doublet of $R$(1, 0) and $R$(1, 1)$^u$ transitions with a
separation of $\Delta v = $ 35~km~s$^{-1}$.

The column density of H$_3^+$ was derived from the observed equivalent
width ($W_\lambda$) using the equation $W_{\lambda} = (8\pi^3\lambda /
3hc) N |\mu|^2$, where $N$ is the column density of H$_3^+$ in the
lower state of the transition.  The values of $|\mu|^2$ listed in
table \ref{tab:lin1} have been used. The velocity-integrated column
densities of individual transitions are summarized in table
\ref{tab:lin1}. For the absorption lines with quality to allow
Gaussian deconvolution, the column density of each velocity component
was calculated and given in table \ref{tab:vel1}.

\section{Discussion}

\subsection{Location of H$_3^+$ Absorbing Clouds}
We discuss below possible identifications of the absorbing clouds,
referring to previous observations at radio and infrared wavelengths
s(figure \ref{fig:h2co}). We will base the following velocity
discussion on GC~IRS~3 spectral lines unless otherwise stated, since
the line profiles are sharper than those of GCS~3-2.

\subsubsection{$-$140~km~s$^{-1}$ to $-$110~km Component}

The $-$140~km~s$^{-1}$ component is attributed to clouds in the
``expanding molecular ring'' \citep{kai72,sco72}. The ``expanding
molecular ring'' is a chain of molecular clouds orbiting around
the nucleus at 200~pc from the Galactic center, and gradually
receding from it. The $-$140~km~s$^{-1}$ component of GC~IRS~3
appears at $-$110~km~s$^{-1}$ in GCS~3-2, which is reasonable because
the ``expanding molecular ring'' appears with a less negative
velocity at positive Galactic longitude. The additional possible
peak at $-$170~km~s$^{-1}$ seen towards GCS~3-2 could be the high
velocity component discussed by \citet{gue81}, \citet{lis93}, and
\citet{yus93}.

\subsubsection{$-$60~km~s$^{-1}$ Component}

A clear match of the peak velocities at $-$60~km~s$^{-1}$ is noted
between the H$_3^+$ lines and the H$_2$CO line observed in absorption
toward Sgr A (Snyder et al. 1969; G\"usten, Downes 1981), and the H
{\sc i} absorption line \citep{lis85}. The cloud velocity is
consistent with the radial velocity of the ``3~kpc arm''. The
``nuclear disk'' rotating at 300~pc from the Galactic center (Sanders,
Wrixon 1973) also shows a similar velocity at the longitude of
GC~IRS~3 (Whiteoak, Gardner 1979). However, we infer that the
contribution of the ``3~kpc arm'' clouds is dominant, since the
component is also seen in GCS~3-2 at the same velocity. This
interpretation is also supported by the absence of a positive velocity
component which is expected from the ``nuclear disk'' for the positive
Galactic longitude of GCS 3-2.

\subsubsection{0~km$^{-1}$ Component}

The 0~km~s$^{-1}$ component is also apparent in the H$_2$CO and H {\sc
i} absorption spectra of Sgr A in figure \ref{fig:h2co}. The
0~km~s$^{-1}$ absorbers are usually attributed to the ``local clouds''
within a few kpc of the solar neighborhood, which is most evident in
the H {\sc i} absorption at 21~cm (e.g., Garwood, Dickey 1989). There
should also be a contribution from the low-velocity dense clouds close
to Sgr~A. Some of them are reported to be within about 50~pc of Sgr~A
by \citet{whi78}.

\subsubsection{+50~km$^{-1}$ Wing}
A candidate for the H$_3^+$ absorption wing at +50~km~s$^{-1}$ is the
``+50~km~s$^{-1}$ clouds,'' a complex of giant molecular clouds within
about 10~pc of the Galactic nucleus (G\"usten et al. 1981; G\"usten,
Henkel 1983). In the model of the large-scale structures in the
central 10~pc of our Galaxy constructed with recent high spatial
resolution radio observations (Coil, Ho 2000; Wright et al. 2001), the
Sgr~A~East non-thermal radio source is impacting the ``+50~km~s$^{-1}$
clouds'' at the far side of the Galactic center. If the H$_3^+$
absorption wing at +50~km~s$^{-1}$ is associated with these giant
molecular clouds, it places GC~IRS~3 beyond Sgr A, which then
jeopardizes the membership of GC~IRS~3 in the central star
cluster. Instead, to account for the absorption feature of CO in the
near-infrared at 4.7~$\mu$m, \citet{geb89} proposed that the
particular cloud occulting GC~IRS~3 could be a part of the
``circumnuclear disk'' delineated by \citet{gue87} in their HCN
map. The ``circumnuclear disk'' is a compact ($\sim$ 3~pc) clumpy
torus, a reservoir supposedly feeding mass to the dynamical center of
our Galaxy, Sgr~A$^\ast$. The +50~km~s$^{-1}$ component is barely seen
in GCS~3-2, which suggests that the absorber of GC~IRS~3 is a compact
local structure at Sgr~A, and thus argues in favor of the
``circumnuclear disk'' origin.

\subsection{Special Characteristic of the H$_3^+$ Number Density}

Lines of sight toward the Galactic center sample both dense and
diffuse clouds along the long pathlength. About 30\% of the visual
extinction toward the Galactic center is believed to arise in dense
clouds, and the rest in diffuse clouds (e.g. Whittet et al. 1997). For
instance, the diffuse and dense cloud extinctions to GC~IRS~3
estimated from the optical depths of hydrocarbon and water ice in the
3~$\mu$m region \citep{chi02} are $A_{V}$ = 36 and $A_{V}$ = 11,
respectively.

The optical depth of H$_3^+$, however, does not scale with the visible
extinction. In contrast to most other molecules, the number density of
H$_3^+$ is constant and independent of the density of the cloud as
long as the number density of H$_2$ relative to that of the destroyer
X of H$_3^+$, $n$(H$_2$)/$n$(X), is constant (figure
\ref{fig:dens1}). A simple analysis using a steady state chemical
kinetics yields a H$_3^+$ number density of $n$(H$_3^+$) =
($\zeta$/$k_{\rm CO}$)[$n$(H$_2$)/$n$(CO)] for a typical dense cloud
where carbon atoms are mostly in the form of CO, and $n$(H$_3^+$) =
($\zeta$/$k_{\rm e}$)[$n$(H$_2$)/$n$(${\rm e}$)] for a typical diffuse
cloud where carbon atoms are in the form of C$^+$ ($\zeta$ is the
cosmic ray ionization rate, and $k_{\rm CO}$ and $k_{\rm e}$ are the
rate constants of the ion-neutral reaction between H$_3^+$ and CO and
of dissociative electron recombination of H$_3^+$, respectively;
McCall et al. 1998a,b). If typical values of $\zeta \sim 3 \times
10^{-17}$ s$^{-1}$, $k_{\rm CO} \sim$ 2 $\times$ 10$^{-9}$ cm$^3$
s$^{-1}$, $k_{\rm e} \sim 5 \times 10^{-7}$ cm$^{3}$ s$^{-1}$ are
used, and [$n$(H$_2$)/$n$(CO)] and [$n$(H$_2$)/$n$(${\rm e}$)] are
both assumed to be 7 $\times$ 10$^3$, we obtain $n$(H$_3^+$) $\sim$ 1
$\times$ 10$^{-4}$ cm$^{-3}$ in dense clouds and $\sim$ 4 $\times$
10$^{-7}$ cm$^{-3}$ in diffuse clouds. These constancies of the number
density make H$_3^+$ nearly a direct indicator of the dimension of the
absorbing clouds. In other words, regardless of how high or low the
cloud density is, the H$_3^+$ column density is simply proportional to
the physical dimension of the cloud along the line of sight. This
simple analysis well explains the observed H$_3^+$ column densities in
dense clouds \citep{mcc99}.

However, a problem appears when this model is applied to diffuse
clouds because of uncertainties of the constants used in the
calculation. Since the observed H$_3^+$ column densities in dense and
diffuse clouds are comparable, the large factor of 250 difference in
$n$(H$_3^+$) leads to the same large difference in the cloud
dimension, which is difficult to accept (McCall et al. 1998a,b;
Geballe et al. 1999; McCall et al. 2002). Currently, at least one of
the three assumed values, [$\zeta$, $k_{\rm e}$, and $n$(${\rm e}$)]
is under suspicion.  For instance, measurements of $k_{\rm e}$ vary by
more than one order depending on the experimental techniques
(e.g. Larsson 2000).
It is likely that the estimate of $n$(H$_3^+$) in
diffuse clouds will be increased by one order of magnitude or so when
the uncertainty in these three parameters is resolved;  
$n$(H$_3^+$) will then give a reasonable cloud dimension.

\subsection{H$_3^+$ in Diffuse Clouds}

The sharp and weak H$_2$CO lines toward Sgr~A at $-$140~km~s$^{-1}$,
$-$60~km~s$^{-1}$ and 0~km~s$^{-1}$, and the weakness of the CO
absorption at these velocities in GC~IRS~3 indicate that the clouds
responsible for those absorptions are diffuse. \citet{geb89} estimated
the CO column density in the cloud at $-$60~km~s$^{-1}$ velocity to be
1 $\times$ 10$^{17}$ cm$^{-2}$ with a cloud temperature of 17~K. If we
use [H$_2$]/[CO] = 7 $\times$ 10$^3$, we obtain $N$(H$_2$) = 7
$\times$ 10$^{20}$ cm$^{-2}$, though this value might be better taken
as a lower limit because of the possible saturation in the CO
fundamental lines.  Such a low hydrogen column density is consistent
with diffuse clouds.

The observed H$_3^+$ column density in each of these velocity
components of GC IRS 3 is in the range $N$(H$_3^+$) = (2.4--5.6)
$\times$ 10$^{14}$ cm$^{-2}$. If the H$_3^+$ number density in diffuse
clouds given above (4 $\times$ 10$^{-7}$ cm$^{-3}$) is assumed, we
obtain unreasonably high cloud dimensions of 200 to 500~pc. The
enigma of H$_3^+$ chemistry in the diffuse interstellar medium needs
to be solved before we discuss the cloud dimension any further.

The large pedestal component of GC IRS 3 is also interpreted to be due
to diffuse clouds, because of the absence of strong CO or H$_2$CO
absorption. A clarification of the nature of this cloud is an
interesting future problem. There exists a possibility that the
decomposition of the observed features into velocity components is not
unique in view of the significant noise in the spectrum and the
limited spectral resolution.

\subsection{H$_3^+$ in Dense Clouds}

The strong and saturated CO absorption toward GC~IRS~3 and H$_2$CO for
Sgr~A at the velocity of +50~km~s$^{-1}$ wing clearly represents dense
clouds (figure \ref{fig:h2co}). It is remarkable that the H$_3^+$
absorption lines starting from the $J$ = 1 levels are weak at this
velocity (figure \ref{fig:spe1}). This must be due to a small
pathlength of a very dense cloud in front of the infrared source. From
the equivalent width of the $R$(1, 1)$^l$ line integrated over +20 to
+60~km~$^{-1}$, $N$(H$_3^+$)$_{J=1}$ = 2.4 $\times 10^{14}$ cm$^{-1}$
is obtained; this corresponds to a dense cloud pathlength of 1.1~pc if
$n$(H$_3^+$) $\sim 1 \times 10^{-4}$ cm$^{-3}$ estimated in subsection
5.2 is assumed. The concentration of the H$_3^+$ in the (3, 3) level
at the velocity would increase the total column density and the cloud
dimension; however, we leave this population not included because we
do not have sufficient knowledge about either the nature of the cloud
that holds H$_3^+$ in the (3, 3) level or the mechanism that populates
the metastable states, which we discuss separately in the next
subsection.

The cloud dimension of the +50~km~s$^{-1}$ wing might not be
incompatible with the ``+50~km~s$^{-1}$ clouds'' interpretation in
terms of the physical parameters presented by \citet{gue83}.
They estimate the hydrogen column density of M-0.02-0.07, the closest
``+50~km~s$^{-1}$ cloud'' to GC~IRS~3, to be $N$(H$_2$) = 7 $\times$
10$^{24}$~cm$^{-2}$, which gives $n$(H$_2$) = 2 $\times$ 10$^{6}$
cm$^{-3}$ if a cloud dimension of 1~pc is assumed. The angular
diameter of M-0.02-0.07 is 2\arcmin~ in \citet{gue83}, which is 5~pc
at the assumed 8~kpc distance. The apparent smaller pathlength
measured by H$_3^+$ could be reconciled if the line of sight toward
GC~IRS~3 is off-centered from the cloud core.

On the other hand the physical model of the ``circumnuclear disk''
proposed by \citet{mar95} gives the radial thickness of the annular
disk to be 0.5~pc with a hydrogen density of $n$(H$_2$) $\simeq$
10$^{6}$ cm$^{-3}$.  The factor 2--3 difference could be filled by the
geometry of GC~IRS~3 and the disk.



\subsection{H$_3^+$ in Metastable State}

The tentative detection of the $R$(3, 3)$^l$ transition at
3.534~$\mu$m for the first time in interstellar space would be a
breakthrough if confirmed by more observations; it introduces a new
dimension in the studies of interstellar H$_3^+$.  The detection of
$R$(3, 3)$^l$ and non-detection of other transitions starting from $J
>$ 1 levels, such as $R$(2, 1)$^u$, $R$(2, 2)$^l$, $Q$(2, 1)$^l$,
$Q$(3, 0) etc., is reasonable since only the (3, 3) rotational level
is metastable; spontaneous emission from this level to lower levels is
forbidden by the ortho--para selection rule and the absence of the (2,
0) rotational level by the Pauli principle, as shown in figure
\ref{fig:pan2} (Pan, Oka 1986).  On the other hand, H$_3^+$ in the (2,
1), (2, 2), (3, 0), (3, 1), and (3, 2) levels decays to lower levels
with lifetimes of 20.4 d, 27.2 d, 3.8 hr, 7.9 hr and 15.8 hr,
respectively (Neale et al. 1996) through centrifugal
distortion-induced rotational transitions initially proposed for
NH$_3$ \citep{oka71}. Since detailed \emph{ab initio} quantum chemical
calculations of H$_3^+$ have been extensively carried out, these
lifetimes are accurate and reliable.  All ortho-H$_3^+$ which are
chemically produced or collisionally pumped into the (4, 3), (3, 0),
and (3, 3) levels accumulate in the (3, 3) level until it is
collisionally deexcited to lower levels.  Similarly, ortho-H$_3^+$
produced in (5, 3), (5, 0), (6, 6), (6, 3), (7, 6), (7, 3), (7, 0)
etc. accumulate in the (6, 6) metastable level. Metastable levels are
also expected for para-H$_3^+$, for which (4, 4) and (5, 5) may be
excessively populated.  H$_3^+$ in the (4, 4) level can spontaneously
decay to the (3, 1) level, but its lifetime is long (11 yr).

H$_3^+$ in metastable levels relaxes to lower levels by collisions
with H$_2$. Unlike NH$_3$ in which ortho $\leftrightarrow$ para
transitions are forbidden for collision-induced transitions
\citep{che69}, H$_3^+$ may relax to lower levels of different spin
modification. This is because a collision of H$_3^+$ with H$_2$ is a
chemical reaction and the scrambling of protons may change
ortho-H$_3^+$ into para-H$_3^+$ or vice versa, although some selection
rules for nuclear spin still remain (Uy et al. 1997; Cordonnier et
al. 2000).

There are two ways that the (3, 3) level could be significantly
populated compared with the lowest $J$ = 1 levels. First, it may be
simply because of high temperature. The (3, 3) level is higher than
the lowest (1, 1) level by 251.23~cm$^{-1}$ (361~K) and the
populations of the two levels become equal at $T$ = 234~K.
Populations in excited non-metastable levels may still be low even at
high temperature because of fast spontaneous emission, if the cloud
density is lower than the critical densities of the levels. The
critical densities are on the order of 10$^4$ cm$^{-3}$ for (2, 1) and
(2, 2) and 10$^6$ cm$^{-3}$ for (3, 0), (3, 1) and (3, 2). Thus our
observation of the $R$(3, 3)$^l$ transition and non-detection of other
transitions starting from $J$ $>$ 1 levels indicate low density. If
the metastable state is positively confirmed to be thermally
populated, then the $R$(3, 3)$^l$ transition and other absorption
lines from metastable levels will make an excellent probe to
efficiently isolate the warm interstellar gas. These absorption lines
are unique in that they are sensitive exclusively to the high
temperature clouds ($T \sim$ 200~K) along the line of sight with
minimum contamination by other cold clouds.

Second, the excess population at (3, 3) could occur at low temperature
with a subtle balance of the rate of collisional deexcitation and
destruction of H$_3^+$. If the former rate, $k_{\rm H}
n$(H$_3^+$)$n$(H$_2$), is much faster than the latter, $k_{\rm X}
n$(H$_3^+$)$n$(X), H$_3^+$ at low temperature will all relax to the
lowest $J$ = 1 levels during its lifetime. The excess population in
(3, 3) could occur only when those rates are comparable, that is,
$k_{\rm H} n$(H$_2$) $\sim$ $k_{\rm X} n$(H$_{\rm X}$). This cannot
happen in dense clouds where CO is the main destroyer of H$_3^+$,
since $k_{\rm H}$ $\sim$ $k_{\rm CO}$ and $n$(H$_2$) $\gg$
$n$(CO). For diffuse clouds this may be possible if
$n$(H$_2$)/$n$(${\rm e}$) is not much larger than $k_{\rm e}$/$k_{\rm
H}$ $\sim$ 250. For this to happen a high electron density and high
$k_{\rm e}$ are preferred. It is interesting to note that the latter
requirement is opposite to what is needed to reconcile the enigma of
the observed high column densities of H$_3^+$ in a diffuse
interstellar medium.

Although our detections of the $R$(3, 3)$^l$ line look fairly
convincing for both GC~IRS~3 and GCS~3-2, we note that both lines are
broad and without a sharp velocity component. In addition, the spectral
line toward GC~IRS~3 is enigmatic in that the five spectral lines
starting from the $J$ = 1 level are not strongly observed at the
velocity of the $R$(3, 3)$^l$ line. A simple model calculation shows
that it is hard to populate the (3, 3) level alone without populating
the (1, 0) and (1, 1) levels. It is intriguing that the line profile of
$R$(3, 3)$^l$ is similar to those of the strongly saturated CO line
both in its wavelength and widths (figure \ref{fig:h2co}). Such CO
lines indicate dense clouds where H$_3^+$ will be collisionally cooled
rapidly to lower levels. We may be sampling low-density warm gas with
a long pathlength surrounding such dense cloud.

The high population in the (3, 3) level makes the problem of the
unexpectedly high abundance in diffuse clouds even harder to
reconcile, since other metastable states may also be significantly
populated. The need of a high value of $k_{\rm e}$ for the second
mechanism to populate (3, 3) is also an interesting twist related to
the enigma. We will attempt to observe more objects and more spectral
lines starting from metastable levels, and to carry out a detailed
analysis of this problem.

\section{Summary}
We have presented an absorption-line survey of H$_3^+$ toward two
Galactic center sources: GCS~3-2 in the Quintuplet cluster and
GC~IRS~3 near Sgr~A$^\ast$. Six H$_3^+$ lines were detected for each
source, of which three were newly detected. In particular, the
absorption of H$_3^+$ originating from the ($J$, $K$) = (3, 3)
metastable state was tentatively detected for the first time in
interstellar space. The observed H$_3^+$ absorption lines show
intriguing line profiles, indicating at least four velocity components
of clouds along the line of sight. The velocities well match those of
H {\sc i}, CO, and H$_2$CO reported earlier in the radio and
infrared. The H$_3^+$ velocity components at 0, $-$60, and $-$110 --
$-$140 km~s$^{-1}$, which well match those of sharp and weak CO as
well as with H$_2$CO, are inferred to be in the diffuse interstellar
medium of intervening spiral arms, while the +50 km~s$^{-1}$ component
in GC~IRS~3 should be associated with the local structure of the
Galactic nucleus. Surprisingly, not much H$_3^+$ is found at +50
km~s$^{-1}$ where very strong and saturated CO absorption is observed
in GC~IRS~3. This shows the marked contrast between CO and H$_3^+$ as
astrophysical probes; the number density of CO is proportional to the
cloud density, while that of H$_3^+$ is independent of the cloud
density.

The tentative detection of the $R$(3, 3)$^l$ line provides
observational evidence for the metastability of the (3, 3) level,
which had been theoretically expected. It suggests that other
metastable $J$ = $K$ levels may also be populated. While the observed
H$_3^+$ spectra reveal the great richness of the sight lines toward
the Galactic center, there are many loose ends in our interpretation
of the spectra. Further studies on the H$_3^+$ spectrum as well as on
CO and H$_2$ spectroscopy toward the Galactic center will be attempted
to further clarify this situation.

\bigskip
We acknowledge all of the staff and crew of the Subaru Telescope and
NAOJ for their valuable assistance in obtaining this data and their
continuous support for the construction of IRCS. We wish to thank an
anonymous referee for useful comments on the manuscript. Special
thanks goes to K. S. Usuda for many inspiring discussions. B. J. M. is
supported by the Miller Institute for Basic Research in
Science. T. O. is supported by NSF grant PHY 00-99442. T. R. G.'s
research is supported by the Gemini Observatory, which is operated by
the Association of Universities for Research in Astronomy, Inc., on
behalf of the international Gemini partnership of Argentina,
Australia, Brazil, Canada, Chile, the United Kingdom and the United
States of America. M. G. is supported by a Japan Society for the
Promotion of Science fellowship. Last, but not least, we wish to
express our deep appreciation to those of Hawaiian ancestry on whose
sacred mountain we are privileged to be guests.

\clearpage
\clearpage
  \begin{table}
  \caption{Major absorption lines of H$_3^+$ $\nu_2$ $ \leftarrow$ 0
  targeted in the survey. }\label{tab:lin1}

\begin{center}
  \begin{tabular}{lccccccc}

\hline \hline
Transition  & $\lambda_{\rm Lab}$     & $|\mu|^2$    & Coverage &\multicolumn{2}{c}{$W_\lambda$}&\multicolumn{2}{c}{$N_{\rm level}$}\\
            &   [$\mu$m]$^\ast$       &  [D$^2$]$^\dag$&       &\multicolumn{2}{c}{[10$^{-5} \mu$m]}&\multicolumn{2}{c}{[10$^{15}$ cm$^{-2}$]}\\
\hline
            &                         &            &          &     GCS~3-2  & GC~IRS3      &   GCS~3-2 & GC~IRS3  \\
\hline
$R$(3, 3)$^u$ \dots& 3.426974         &   0.0071   &   No     &              &              &           &          \\
$R$(3, 3)$^l$ \dots& 3.533666         &   0.0191   &   Yes    &    2.8       &   3.2        &  1.0      &  1.1     \\
$R$(2, 1)$^u$ \dots& 3.538424         &   0.0182   &   Yes    &  $\cdots$    & $\cdots$     &$\cdots$   &$\cdots$  \\
$R$(2, 2)$^u$ \dots& 3.542158         &   0.0094   &   Yes    &  $\cdots$    & $\cdots$     &$\cdots$   &$\cdots$  \\
$R$(2, 1)$^l$ \dots& 3.615924         &   0.0044   &   No     &              &              &           &          \\
$R$(2, 2)$^l$ \dots& 3.620473         &   0.0177   &   No     &              &              &           &          \\
$R$(1, 1)$^u$ \dots& 3.668083         &   0.0158   &   Yes    &  11$^\ddag$  &  8.2$^\ddag$ & 2.5$^\S$  & 1.9$^\S$ \\
$R$(1, 0)     \dots& 3.668516         &   0.0259   &   Yes    &  11$^\ddag$  &  8.2$^\ddag$ & 1.2$^\S$  & 0.94$^\S$\\
$R$(1, 1)$^l$ \dots& 3.715479         &   0.0141   &   Yes    &  5.1$^{||}$  &  4.8$^{||}$  &   2.3     &  2.2     \\
$Q$(3, 3)     \dots& 3.903967         &   0.0065   &   No     &              &              &           &          \\
$Q$(2, 2)     \dots& 3.914406         &   0.0086   &   Yes    &  $\cdots$    & $\cdots$     & $\cdots$  &$\cdots$  \\
$Q$(2, 1)$^u$ \dots& 3.916979         &   0.0043   &   Yes    &  $\cdots$    & $\cdots$     & $\cdots$  &$\cdots$  \\
$Q$(1, 1)     \dots& 3.928625         &   0.0128   &   Yes    &    4.5       &   5.1        &   2.1     &  2.5     \\
$Q$(1, 0)     \dots& 3.953000         &   0.0254   &   Yes    &    5.2       &   5.0        &   1.3     &  1.2     \\
$Q$(2, 1)$^l$ \dots& 3.971073         &   0.0167   &   Yes    & $\cdots$     & $\cdots$     & $\cdots$  &$\cdots$  \\
$Q$(3, 0)     \dots& 3.985528         &   0.0249   &   Yes    & $\cdots$     & $\cdots$     & $\cdots$  &$\cdots$  \\
$Q$(3, 1)$^u$ \dots& 3.987028         &   0.0217   &   Yes    & $\cdots$     & $\cdots$     & $\cdots$  &$\cdots$  \\
$P$(1, 1)     \dots& 4.069524         &   0.0086   &   No     &              &              &           &          \\
\hline
$N_{\rm total}$ [10$^{15}$ cm$^{-2}$]&&            &          &              &              &   4.6     &  4.4     \\
\hline
    \end{tabular}

  \end{center}
Note. The laboratory data are from
  \citet{mck98} and \citet{lin01}. The velocity-integrated equivalent
  widths and the column densities are summarized for the detected lines.

$^\ast$  \citet{lin01}.\\
$^\dag$  J. K. G. Watson, private communication.\\
$^\ddag$ Total equivalent width of $R$(1, 1)$^u$ and $R$(1, 0).\\
$^\S$    Column density of the each level is derived on the assumption that
         the populations in the ortho- (1, 0) and para- (1, 1) levels are divided into about 2 : 1,
         which is implied by $R$(1, 1)$^l$, $Q$(1, 1) and $Q$(1, 0) transitions.  \\
$^{||}$  Equivalent width of $R$(1, 1)$^l$ is calculated from the Gaussian decomposition parameters.
         For other transitions $W_\lambda$ are obtained by directly integrating the absorption features. \\

\end{table}

\clearpage
\begin{table}
  \caption{Summary of observations.}\label{tab:log1}

\begin{center}
  \begin{tabular}{lcccccccccc}
\hline \hline  
Name                &  R.A.        &  Dec.       &   $l$    & $b$      & $L$     &  Exposure &\multicolumn{2}{c}{Grating$^\ast$}&\multicolumn{2}{c}{Standard}\\
                    &     (J2000)  &     (J2000) & [\degree]& [\degree]&         &       [s] & ECH   & XDP                   &   Name          & Spe.         \\
\hline

GCS~3-2$^\dag$\dots & 17:46:14.8   & $-$28:49:41 &  +0.16   &$-$0.06   & 2.7     &  120      & 8350  &   6100  & HR~7528 & B9.5~IV\\
                    &              &             &          &          &         &  240      & 4400  &   5200  & HR~7528 & B9.5~IV\\
                    &              &             &          &          &         &           &       &         &                 \\
GC~IRS~3$^\ddag$\dots& 17:45:39.9   & $-$29:00:24 &$-$0.06   &$-$0.04   & 5.3     &  360      & 8350  &   6100  & HR~7924 ($\alpha$ Cyg) & A2~Iae\\
                    &              &             &          &          &         &  1440     & 4400  &   5200  & HR~7924 ($\alpha$ Cyg) & A2~Iae\\

\hline

    \end{tabular}
\end{center}
$^\ast$  ``ECH'' and ``XDP'' denote the angle of echelle and cross-dispersing gratings in the instrumental unit. \\
$^\dag$  Coordinate and magnitude are from \citet{nag90}.\\
$^\ddag$ Coordinate and magnitude are from \citet{blu96}.\\

\end{table}
\clearpage
\begin{table}
  \caption{Velocity-resolved components of $R$(1, 1)$^l$, $Q$(1,
  1) and $Q$(1, 0).} \label{tab:vel1}

\begin{center}
  \begin{tabular}{lccccccccc}

\hline \hline
Name          & $v_{\rm LSR}$ &  FWHM          & \multicolumn{3}{c}{$W_{\lambda}$}   & \multicolumn{3}{c}{$N_{\rm level}$}      & $N_{\rm total}    $       \\
              & [km~s$^{-1}$] &  [km~s$^{-1}$] & \multicolumn{3}{c}{[10$^{-5}~\mu$m]}& \multicolumn{3}{c}{[10$^{14}$ cm$^{-2}$]}& [10$^{14}$ cm$^{-2}$]$^\ast$ \\
\hline
              &               &                & $R$(1, 1)$^l$&$Q$(1, 1)& $Q$(1, 0)  & $R$(1, 1)$^l$&$Q$(1, 1)& $Q$(1, 0)       &                \\

\hline

GCS~3-2\dots  &    $-$165     &     15   &      0.15 &     $-$           &   $-$     &  0.7  &     $-$           &   $-$     &  1.3\\
              &    $-$107     &     50   &      1.8  &     2.0           &   1.2     &  8.3  &     9.4           &   2.9     &  16 \\
              &     $-$51     &     38   &      1.9  &     1.7           &   2.1     &  8.6  &     7.9           &   5.1     &  17  \\
              &      $-$2     &     31   &      1.2  &     1.5           &   1.7     &  5.6  &     7.0           &   4.1     &  11  \\
              &               &          &           &                   &           &       &                   &           &      \\
GC~IRS~3\dots &    $-$140     &     22   &      0.64 &     $-$           &   0.30    &  2.9  &     $-$           &   0.51    &  5.6 \\
              &     $-$56     &     15   &      0.27 &     $-$           &   0.46    &  1.2  &     $-$           &   1.1     &  2.4 \\
              &     $-$23~$^\dag$&116    &      3.7  &     $-$           &   3.0     &  17   &     $-$           &   7.2     &  32 \\
              &     $-$5      &     12   &      0.50 &     $-$           &   0.73    &  2.3  &     $-$           &   1.7     &  4.3 \\
\hline
 \end{tabular}
\end{center}
 
$^\ast$ Based on the velocity independent conversion factor of $R$(1, 1)$^l$ to total column density derived from table 1.\\  
$^\dag$ Pedestal component.

\end{table}

\clearpage
\begin{figure}
  \begin{center}
  \end{center}

  \caption{Wavelength coverage of the IRCS echelle mode in the
     3~$\mu$m region. The entire $L$-band (2.84--4.18~$\mu$m) is
     covered by six standard settings (LA$^-$, LA$_0$, LA$^+$, LB$^-$,
     LB$_0$, LB$^+$; shown in green and blue rectangles), but 70\% is
     covered by three settings. ``ECH'' and ``XDP'' denote the angles
     of the echelle and the cross-disperser gratings. The grating
     angle can be seamlessly modified so as to be best suited to
     individual projects. The two settings which we employed for the
     observation are shown by the bold-line enclosures. The positions
     of the major H$_3^+$ lines are indicated by sticks with the
     heights representing the relative intensities calculated with $T$
     = 300~K. The atmospheric transmission curve is overlaid to
     illustrate the region where the interference of the telluric
     atmospheric absorption is severe.}  \label{fig:cov1}

\end{figure}

\begin{figure}
  \begin{center}
  \end{center}

   \caption{Total detected H$_3^+$ absorption spectra toward GCS~3-2
     (left column) and GC~IRS~3 (right column). Kinematic components
     are indicated in the LSR velocity by dotted lines (See figure
     \ref{fig:dec1}). Note that the 3.668~$\mu$m line profile appears
     to be different from the others because it consists of a close
     doublet of $R$(1, 0) and $R$(1, 1)$^u$ separated by $\Delta v$ =
     35~km~s$^{-1}$. The feature is lined up with other absorption
     lines at $R$(1, 0) 3.6685~$\mu$m. The model atmospheric
     transmission curves are shown to discriminate the genuine
     detections from the possible artifacts of the poor cancellation
     of the telluric absorption lines.}  \label{fig:spe1}
\end{figure}

\begin{figure}
  \begin{center}
   \end{center}

     \caption{Same with figure \ref{fig:spe1}, but for negative
     detections. The periodic patterns seen at 3.97--3.99~$\mu$m are
     instrumental fringes left unprocessed. Note that the dips at
     3.534~$\mu$m in the bottom panels are $R$(3, 3)$^l$ positively
     detected.}\label{fig:neg2}

\end{figure}

\begin{figure}
  \begin{center}
  \end{center}

     \caption{Gaussian decomposition of the kinematic components for
     GCS~3-2 (top) and GC~IRS~3 (bottom). At least four components are
     found in ortho- $Q$(1, 0) and para- $R$(1, 1)$^l$ and $Q$(1, 1)
     H$_3^+$ absorption lines at common velocities. The residual of
     the fitting is shown at the bottom of each panel. The
     decomposition parameters are summarized in table
     \ref{tab:vel1}. }\label{fig:dec1}
\end{figure}

\begin{figure}
  \begin{center}
  \end{center}

     \caption{Comparison of the line profiles of H$_3^+$ at $R$(1,
1)$^l$ and $R$(3, 3)$^l$ along with those of other species toward the
same objects. GCS~3-2 is on the left, and GC~IRS~3 on the right. The
absorption spectra from previous studies are shown in arbitrary
units. The dotted lines represent the velocity components of the
$R$(1, 1)$^l$ transition in the previous figure.  Three velocity
components in GC~IRS~3 at $-$140~km~s$^{-1}$, $-$60~km~s$^{-1}$, and
0~km~s$^{-1}$ match well with CO $R$(2), H$_2$CO and H {\sc i}
absorption, indicating the same absorbing clouds. The deep absorptions
of CO and H$_2$CO at +50~km~s$^{-1}$ toward the Galactic center do not
correspond to strong components in the H$_3^+$ spectrum, which might
be accounted for by the compact structure of the absorbing cloud. Note
the considerable difference in the line shape of $R$(3, 3)$^l$ from
$R$(1, 1)$^l$ for both GCS~3-2 and GC~IRS~3. The line profile of the
$R$(3, 3)$^l$ in GC~IRS~3 is more or less similar to that of CO
absorption at 4.7~$\mu$m.}
\label{fig:h2co}

\end{figure}

\begin{figure}
  \begin{center}
  \end{center}

     \caption{Schematic of the number density variation of H$_3^+$
     with respect to hydrogen along with that of major molecules and
     ions in the interstellar medium. The number density of H$_3^+$ is
     independent of $n$(H), in contrast with the other molecules. We
     define ``dense'' and ``diffuse'' clouds by the form of carbon
     atoms. The form of carbon is critical, since it defines the
     destruction mechanism of H$_3^+$, and hence the number density of
     the molecular ions.}  \label{fig:dens1}

\end{figure}

\begin{figure}
  \begin{center}
  \end{center}

   \caption{ Rotational energy levels of H$_3^+$ in the ground state
    and in the $\nu_2$ vibrationally excited state.  H$_3^+$ has two
    vibrational modes of which only $\nu_2$ is infrared active, and
    targeted in our observation.  $G$ = $| k - l_2|$ is used instead
    of $K$ in the diagram, where $l_2$ is a vibrational angular
    momentum of $\nu_2$. $G$ is a better quantum number than $K$ where
    $l_2 \ne 0$ (see McCall 2001 for details). The letters $u$ and $l$
    marked in the $\nu_2$ state denote the levels of the same $G$
    number, but with different combinations of $k$ and $l_2$. The
    major transitions of $\nu_2$ $\leftarrow$ 0 that appear in the 3
    to 4~$\mu$m regions are indicated by the vertical connecting
    arrows with labels. Note that for the ground vibrational state,
    the $J$ = 2$n$ and $G$ = 0 levels are forbidden by the Pauli
    principle (shown in dotted bars).  The lowest levels of ortho-
    ($J$, $G$) = (1, 0) and para- ($J$, $G$) = (1, 1) H$_3^+$ are
    marked with thick solid bars.  The selection rules allow only
    radiative relaxation between two levels that satisfy $\Delta J =
    0,~\pm 1$ and $\Delta G = \pm 3$. The spontaneous transitions are
    indicated with connecting lines of energy levels in the ground
    state. The transition between $G = 2\leftrightarrow 1$ can be
    understood to be $G = \pm 2\leftrightarrow \mp 1$. The ``$+$'' and
    ``$-$'' signs in the ground state denote the parity of the energy
    levels. The parity rule, $+$ $\leftrightarrow$ $-$, is
    automatically satisfied by $\Delta G = \pm 3$.  The ($J$, $G$) =
    (3, 3) and (5, 5) are disconnected from any lower levels by the
    selection rule, making the two levels metastable (Pan, Oka 1986).
    The (4, 4) level can relax to (3, 1), but has a long lifetime.

} \label{fig:pan2}

\end{figure}
\setcounter{figure}{0}
\clearpage
\begin{figure}
  \begin{center}
     \FigureFile(90mm,240mm){set1_h3p.ps}
  \end{center}
   \caption{} \label{fig:cov1}
\end{figure}

\clearpage
\begin{figure}
  \begin{center}
     \FigureFile(170mm,240mm){h3p2.ps}
  \end{center}

   \caption{} \label{fig:spe1}
\end{figure}

\clearpage
\begin{figure}
  \begin{center}
    \FigureFile(170mm,240mm){neg2.ps}
   \end{center}

     \caption{}\label{fig:neg2}

\end{figure}

\clearpage
\begin{figure}
  \begin{center}
     \FigureFile(115mm,220mm){ort1.ps}
  \end{center}

     \caption{ }\label{fig:dec1}
\end{figure}

\clearpage
\begin{figure}
  \begin{center}
     \FigureFile(170mm,240mm){irs3h1.ps}
  \end{center}

     \caption{}
      \label{fig:h2co}

\end{figure}

\clearpage
\begin{figure}
  \begin{center}
    \FigureFile(150mm,240mm){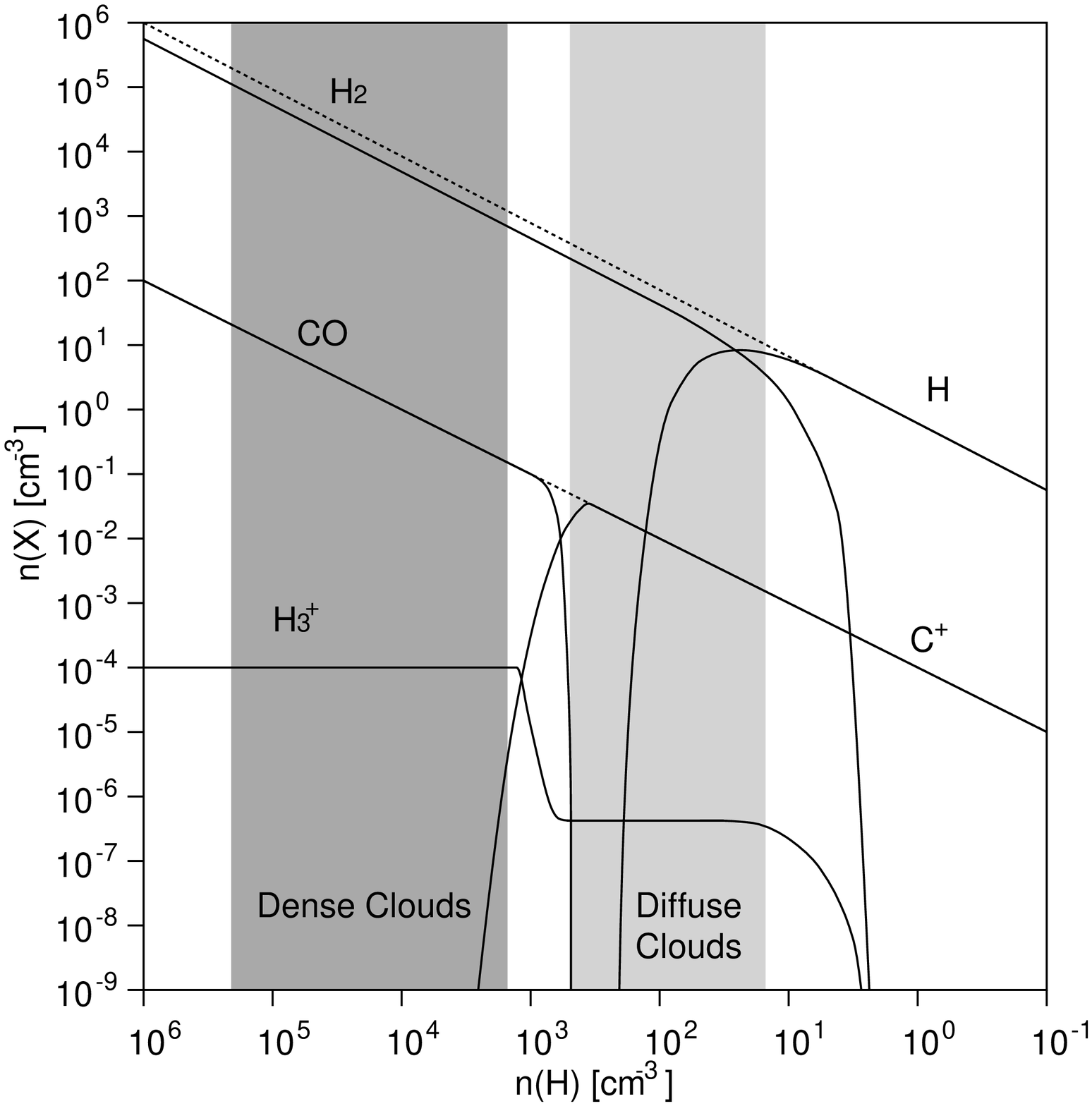}
  \end{center}

     \caption{} \label{fig:dens1}

\end{figure}

\clearpage
\begin{figure}
  \begin{center}
     \FigureFile(115mm,210mm){pan2.ps}
     \clearpage
  \end{center}

   \caption{} \label{fig:pan2}

\end{figure}

\end{document}